# Infrared Reflectance Spectrum of BN Calculated from First Principles

Yongqing Cai, Litong Zhang, Qingfeng Zeng, Laifei Cheng, Yongdong Xu

National Key Laboratory of Thermostructure Composite Materials,

Northwestern Polytechnical University, Xi'an, Shannxi 710072, PR China

ABSTRACT

Using the linear response theory, the vibrational and dielectric properties are calculated for c-BN, w-BN and h-BN. Calculations of the zone-center optical-mode frequencies (including LO-TO splittings) are reported. All optic modes are identified and agreement of theory with experiment is excellent. The static dielectric tensor is decomposed into contributions arising from individual infrared-active phonon modes. It is found that all of the structures have a smaller lattice dielectric constant than that of electronic contribution. Finally, the infrared reflectance spectrums are presented. Our theoretical results indicate that *w*-BN shows a similar reflectivity spectrum as *c*-BN. It is difficult to tell the wurtzite structure from the zinc blende phase by IR spectroscopy.

Keywords: A. BN; C. Infrared reflectance spectrum; E. First principles

## Ⅰ. INTRODUCTION

Boron nitride is one of the most important material among the Ⅲ-Ⅴcompounds due to its chemical, mechanical and electronic properties. High bulk modulus, high thermal conductivity and low dielectric have seen *c*-BN used for protective coating films and modern microelectronic devices at high temperature. Being a very good electrical insulator with good thermal conductivity and stability, *h*-BN has been widely used in nuclear energy and vacuum technology [1].

Boron nitride is known to have three main polytypes: the cubic phase (*c*-BN) with zinc blende structure is the thermodynamically stable phase under ambient conditions, the wurtzite structure (*w*-BN) is mestastable above a pressure of 10Gpa [2], which resembles hexagonal diamond, the hexagonal structure of BN (*h*-BN) is similar to the structure of graphite. Infrared reflectivity (IR) measurement may be considered a powerful tool for the quantitative analysis of these structures. Geick et al. [3] investigated the reflection and transmission spectrum in the spectral range from 400-50000 $cm^{-1}$ and two *h*-BN structures with different stacking sequences of the hexagonal BN layers were discussed. P.J. Gielisse et al. [4] recorded the infrared reflection spectrum of cubic boron nitride between 400-1800 $cm^{-1}$. However, as far as we know, there has been no these experiments on the wurtzite modification.

First-principles calculations based on the density-functional theory complement experiment and are useful for predicting the ground-state properties of materials using no experimental values. Extensive theoretical studies have been performed on the electronic and structural properties of BN [5,6]. For dielectric properties, the static dielectric constants of the three polytypes were calculated with the ultrasoft pseudopotential method and the linear response approach based on density-functional perturbation theory [7]. However, to our best knowledge, there has been no theoretical study on the infrared reflectance spectrum for the three modifications of BN (*c*-BN, *w*-BN and *h*-BN).

In this work, we present the results of first-principles calculations on the vibrational and dielectric properties for *c*-BN, *w*-BN and *h*-BN through linear response approach. The calculated infrared reflectance spectrums are reported. In Sec. Ⅱ， we briefly describe the technical aspects of our calculations. Sec.Ⅲ presents the results, including the structural relaxations, the phonon normal modes, the dielectric tensors and the theoretical IR reflectance spectrums. Sec.Ⅳ concludes the paper.

## Ⅱ. DETAILS OF FIRST-PRINCIPLES CALCULATIONS

The calculations are carried using the CASTEP [8,9] code with Norm-conserving pseudopotentials [10]. We employ the Ceperley-Alder [11] local density functional potential as parameterized by Vosko et al [12]. The 2s and 2p semicore shells are included in the valence for both N and B. The kinetic energy cutoff for the plane waves is 550 ev. Integrals over the Brillouin Zone are approximated by sums on (10×10×10), (9×9×6 ) and (9×9×4) Monkhorst-Pack k-point meshes [13] for *c*-BN, *w*-BN and *h*-BN, respectively.

The first-principles investigation of vibrational and dielectric properties are performed within linear response theory [14]. $\Gamma$-phonon frequencies and dielectric tensors are computed as second-order derivatives of the total energy with respect to atomic displacements or an external electric field. Technical details can be found in [15,16]. To obtain LO/TO splitting characteristic for the three polytypes we introduced to the dynamical matrix the non-analytical term proposed by Pick [17].

## Ⅲ Results and discussion

The calculation starts with a optimization of the lattice constants and the internal coordinates to obtain the minimal energy structural model. Both the lattice constants and the internal coordinates are relaxed by calculating the ab initio forces on the ions until the absolute values of the forces are converged to less than 0.01 ev/Å. For the *c*-BN and *h*-BN case the minimum energy configurations are determined by relaxing the lattice constants only, whereas in the w-BN case the energy is related to both the lattice constants and an internal parameter *u*, which characterizes the relative length of the B-N bonds parallel to the *c* axis.

The optimized structural parameters are listed in Table 1. It can readily be seen that there is excellent agreement between our results and previous experiment except for the lattice constant of *h*-BN. Our calculated value is 6.378 Å whereas the corresponding reported value in Ref. 20 was 6.660 Å. The relatively large discrepancy of the lattice constant *c* is due to the Van-der-Waals interaction between the sheets which is not properly described neither in the LDA nor the GGA [21]. It should be noted that an overall underestimation of the lattice constants in comparison with experimental results is typical of LDA calculations.

Since the primitive unit cells of BN structures have 2 atoms for *c*-BN, 4 atoms for both *w*-BN and *h*-BN, the corresponding numbers of vibrational modes are 6 and 12, respectively.

Our group theoretical analyses indicate that the optical modes at the $\Gamma$ point can be decomposed as

$$\Gamma_{vib}(cubic) = T_2(R, IR)$$

$$\Gamma_{vib}(wurtzite) = A_1(R, IR) + E_1(R, IR) + 2E_2(R) + 2B_1(silent)$$

$$\Gamma_{vib}(hexagonal) = A_{2u}(IR) + E_{1u}(IR) + 2E_{2g}(R) + 2B_{1g}(silent)$$

For *c*-BN, there is a bond stretching $T_2$ mode at the center of Brillouin zone, which is active in both Raman and IR spectroscopy. In the case of the wurtzite and hexagonal structures the IR modes group into modes with displacements either in the *x,y* plane or along the *z* direction. The $E_1$ and $E_{1u}$ modes have displacement patterns in the *x,y* plane, whereas $A_1$ and $A_{2u}$ modes have displacements along z. Due to inversion symmetry, the IR and Raman modes are mutually exclusive for *h*-BN.

Values of the calculated optical phonon frequencies are listed in Table Ⅱ. They are compared with experimental data from first-order Raman or IR spectroscopy [2,22,23] as well as results of previous first principles calculation using ultrasoft pseudopotentials method and the linear response approach [7]. Our calculation presents a root mean square absolute deviation of 21 $cm^{-1}$, and a rms relative deviation of 6.5% with respect to the measurements. The "missing" $E_{2g}$ mode in Ref.2 is identified, which was previously believed in the low frequency region of the spectrum due to the small Van-der-Waals forces between adjacent flat $B_3N_3$ hexagons layers. No experimental data are available for w-BN, however, the agreement of our calculation with other theoretical investigation for the modes at the $\Gamma$ point is reasonable [7].

Crystal symmetry makes the dielectric tensors be composed of some independent components. In the cubic phase, $\varepsilon$ is diagonal in the Cartesian frame with $\varepsilon_{xx} = \varepsilon_{yy} = \varepsilon_{zz}$. For *w*-BN and *h*-BN, the dielectric tensors are diagonal and have two independent components $\varepsilon_{\parallel}$ and $\varepsilon_{\perp}$ along and perpendicular to the c axis, respectively. Our results for the dielectric properties of the three BN modifications are presented in Table Ⅲ. For *c*-BN and *w*-BN an overall excellent agreement between the calculation and experiment is reached. In the case of *h*-BN, there are relatively large deviations concerning the perpendicular components. Our calculated values are 2.95 for electric dielectric constant and 3.57 for static dielectric constant whereas the corresponding reported values in Ref. 3 were 4.1 and 5.06. However, our results are consistent with Ref. 7 and Ref. 26. The differences in comparison with experimental values may be for two reasons: (a) DFT in LDA or GGA is not appropriate to describe the weak interactions in *h*-BN as mentioned above. (b) Poor quality of *h*-BN samples used in the IR reflectivity measurements in Ref. 3, as we discuss below.

The dielectric constant of a material as a function of frequency $\omega$ is given by [27]

$$\varepsilon(\omega) = \varepsilon_{\infty} + \varepsilon_{\infty} \sum_{m} \frac{\omega_{LO,m}^{2} - \omega_{TO,m}^{2}}{\omega_{TO,m}^{2} - \omega^{2} + i\gamma\omega} \qquad (1)$$

where $\varepsilon_{\infty}$ is the electronic contribution to $\varepsilon$, the sum is over IR-active phonon modes *m*, $\gamma$ is the damping coefficients, $\omega_{TO,m}$ and $\omega_{LO,m}$ are their transverse optic (TO) and longitudinal optic (LO) mode frequencies, respectively. From the zero-frequency limit of Eq. (1) $\varepsilon_{0}$ can be separated into contributions arising from purely electronic screening $\varepsilon_{\infty}$ and IR-active phonon modes according to

$$\varepsilon_{0} = \varepsilon_{\infty} + \varepsilon_{\infty} \sum_{m} \frac{\omega_{LO,m}^{2} - \omega_{TO,m}^{2}}{\omega_{TO,m}^{2}} \qquad (2)$$

The IR active mode frequencies and corresponding contributions to the dielectric tensor are displayed in Table Ⅳ. It can be seen that for both polarized directions of the three polytypes only one IR active mode makes contribution to the static dielectric constant. The zinc-blende and wurtzite structures which differ only in the stacking order of double layers of atoms have almost the same lattice dielectric constant, with $\varepsilon_{\parallel}^{lat} = \varepsilon_{\perp}^{lat}$ =2.38 for *c*-BN and $\varepsilon_{\parallel}^{lat}$ =2.58 and $\varepsilon_{\perp}^{lat}$ =2.16 for *w*-BN. For all of the modifications contributions from the electronic polarizations are more than two times larger than that from lattice vibrations.

The infrared reflectivity can be obtained from complex dielectric response $\varepsilon(\omega)$ via

$$R(\omega) = \left| \frac{\sqrt{\varepsilon(\omega)} - 1}{\sqrt{\varepsilon(\omega)} + 1} \right|^2 \qquad (3)$$

In Fig. 1 and Fig. 2 we present our predicted reflectivity $R(\omega)$ spectrum of *c*-BN and *h*-BN, compared with experimental spectrum measured in Ref. 4 and Ref. 3. The damping constant has been chosen for each mode to be 3% of the frequency. It should be noted that two reststrahlen bands in the reflectivity and two active modes for each direction of polarization were found in Ref. 3 in their infrared investigations on *h*-BN. The group theoretical analysis shows that only one mode corresponding to one reststrahlen band for each direction ($A_{2u}$ for $E \parallel c$ and $E_{1u}$ for $E \perp c$) should be expected. The weaker modes in the infrared spectra, according to the analysis of the authors, may probably due to the misorientation in their polycrystalline sample, which may also lead to some uncertainty of the result of the dielectric constant measurement. In Fig. 3 we display our calculated reflectivity $R(\omega)$ spectrum for *w*-BN. Our theoretical results indicate that *w*-BN shows a similar reflectivity spectrum as *c*-BN. It is difficult to distinguish the wurtzite structure from the zinc blende phase by IR spectroscopy. No IR spectrum has been reported for comparison, however, it is highly desirable to obtain polarized single-crystal IR data to confirm our predictions.

## Ⅳ. Conclusion

We have presented a first-principles study of the vibrational and dielectric properties for *c*-BN, *w*-BN and *h*-BN. Firstly, the structural parameters are optimized to obtain the minimal energy structural model and an overall agreement has been reached between our calculated results and previous experiments except for lattice constant *c* of *h*-BN. The vibrational frequencies at the center of Brillouin zone are evaluated. Then the dielectric tensors have been obtained. The mode contributions to the static dielectric tensors are also presented. We find that all of the structures have a smaller lattice dielectric constant than that of electronic contribution. Finally, the infrared reflectance spectrums are reported.

## Acknowledgement

The authors acknowledge support from the Natural Science Foundation of China (Contract No. 90405015), from the National Young Elitists Foundation (Contract No. 50425208) and from the Program for Changjiang Scholars and Innovative Research Team in university (PCSIRT).

## Reference

[1] L. Liu, Y. P. Feng, and Z. X. Shen, Phys. Rev. B 68,104102 (2003).

[2] V.L. Solozhenko, J. Hard Mater. 6, 51 (1995).

[3] R. Geick, C.H. Perry, and G. Ruppercht, Phys. Rev. 146, 543 (1966).

[4] P.J. Gielisse, S.S. Mitra, J.N. Plendl, R.D. Griffis, L.C. Mansur, R. Marshall, and E.A. Pascoe, Phys. Rev. 155, 1039 (1967).

[5] G. Cappellini, G. Satta, M. Palummo, and G. Onida, Phys. Rev. B 64, 035104 (2001).

[6] J. B. MacNaughton et al., Phys. Rev. B 72, 195113 (2005).

[7] Nobuko Ohba, Kazutoshi Miwa, Naoyuki Nagasako, and Atsuo Fukumoto, Phys. Rev. B 63, 115207 (2001).

[8] MD Segall, PJD Lindan, MJ Probert, CJ Pickard, PJ Hasnip, SJ Clark, and MC Payne, J. Phys. Condens. Matter 14, 2717 (2002).

[9] Ackland, G.J.; Warren, M.C.; Clark, S. J., J. Phys.: Cond. Matt. 9, 7861 (1997).

[10] D. R. Hamann, M. Schlüter and C. Chiang, Phys. Rev. Lett. 43, 1494 (1979).

[11] D. M. Ceperley and B. J. Alder, Phys. Rev. Lett. 45, 566 (1980).

[12] S. H. Vosko, L. Wilk and M. Nusair, Can. J. Phys. 58, 1200 (1980).

[13] H. J. Monkhorst and J. D. Pack, Phys. Rev. B 13, 5188 (1976).

[14] S. Baroni, S. de Gironcoli, A. Dal Corso, and P. Giannozzi, Rev. Mod. Phys. 73, 515 (2001).

[15] X. Gonze, Phys. Rev. B 55, 10 337 (1997).

[16] X. Gonze and C. Lee, Phys. Rev. B 55, 10 355 (1997).

[17] R. Pick, M. H. Kohen, and R. M. Martin, Phys. Rev. B 1, 910 (1970).

[18] R.H. Wentorf, Jr., J. Chem. Phys. 26, 956 (1957).

[19] T. Soma, S. Sawaoka, and S. Saito, Mater. Res. Bull. 9, 755 (1974).

[20] V.L. Solozhenko, G. Will, and F. Elf, Solid State Commun. 96, 1 (1995).

[21] L. Wirtz and A. Rubio, Solid State Commun. 131, 141 (2004).

[22] O. Brafman, G. Lengyel, and S.S. Mitra, Solid State Commun. 6, 523 (1968).

[23] T. Kuzuba, K. Era, T. Ishii, and T. Sato, Solid State Commun. 25, 863 (1978).

[24] M.I. Eremets, et al., Phys. Rev. B 52, 8854 (1995).

[25] K. Karch and F. Bechstedt, Phys. Rev. B 56, 7404 (1997).

[26] Y.-N. Xu and W.Y. Ching, Phys. Rev. B 44, 7787 (1991).

[27] V. Železný, et al. Phys. Rev. B 66, 224303 (2002).

TABLE Ⅰ. Calculated and experimental structural parameters for the three polytypes of BN: lattice constant *a*,*c* (Å), internal parameter *u*, and inequivalent atomic positions.

| a | c | u | cation(B) | anion(N) | |
|---|---|---|---|---|---|
| | | *c*-BN | | | |
| 3.592 | | | 0,0,0 | 1/4,1/4,1/4 | present |
| 3.615 | | | | | expt.[a] |
| | | *w*-BN | | | |
| 2.531 | 4.189 | 0.3746 | 1/3,2/3,0 | 1/3,2/3,u | present |
| 2.553 | 4.228 | | | | expt.[b] |
| | | *h*-BN | | | |
| 2.491 | 6.378 | | 1/3,2/3,1/4 | 2/3,1/3,1/4 | present |
| 2.504 | 6.660 | | | | expt.[c] |

[a] Reference 18.

[b] Reference 19.

[c] Reference 20.

TABLE Ⅱ.The optical phonon frequencies (in $cm^{-1}$) at the Γ point of the three polytypes of BN. Two numbers in a row correspond to TO/LO frequencies.

| Material | Mode | present | expt.[a] | expt.[b] | expt.[c] | calc.[d] |
|---|---|---|---|---|---|---|
| *c*-BN | $T_2$ | 1027/1269 | 1056/1304 | | | 1062/1295 |
| *w*-BN | $A_1$ | 1001/1249 | | | | 1043/1280 |
| | $E_1$ | 1041/1267 | | | | 1075/1293 |
| | $E_2$ | 476 | | | | 475 |
| | $E_2$ | 938 | | | | 979 |
| *h*-BN | $A_{2u}$ | 746/819 | | 783/828 | | 754/823 |
| | $E_{1u}$ | 1372/1610 | | 1367/1610 | | 1382/1614 |
| | $E_{2g}$ | 59 | | | 52 | 50 |
| | $E_{2g}$ | 1372 | | 1370 | 1366 | 1382 |

[a] Reference 22.

[b] Reference 3.

[c] Reference 23.

[d] Reference 7.

TABLE Ⅲ. Calculated dielectric properties for three polytypes of BN: macroscopic dielectric constant $\varepsilon_{\infty}$, static dielectric constant $\varepsilon_{0}$.

| Material | $\varepsilon_{\perp}^{\infty}$ | $\varepsilon_{\parallel}^{\infty}$ | $\varepsilon_{\perp}^{0}$ | $\varepsilon_{\parallel}^{0}$ | |
|---|---|---|---|---|---|
| *c*-BN | 4.52 | 4.52 | 6.93 | 6.93 | present |
| | 4.46 | 4.46 | 6.8 | 6.8 | expt.[a] |
| | 4.54 | 4.54 | 6.74 | 6.74 | calc.[b] |
| *w*-BN | 4.49 | 4.64 | 6.67 | 7.25 | present |
| | 4.50 | 4.67 | | | calc.[c] |
| *h*-BN | 4.87 | 2.95 | 6.71 | 3.57 | present |
| | 4.95 | 4.1 | 6.85 | 5.06 | expt.[d] |
| | 4.85 | 2.84 | 6.61 | 3.38 | calc.[b] |

[a] Reference 24.
[b] Reference 7.
[c] Reference 25.
[d] Reference 3.

TABLE Ⅳ. LO/TO splittings and mode contribution to the component of dielectric tensor of *c*-BN, *w*-BN and *h*-BN. The first column (E=0) is for no electric field, the second is for the field lying in the plane (E∥a-b), the third is for E∥c. A component of lattice dielectric constant parallel (perpendicular) to c-axis is donated $\varepsilon_{\parallel}^{lat}$ ($\varepsilon_{\perp}^{lat}$). $\Delta\omega^2 = \omega_{\mathrm{LO}}^2 - \omega_{\mathrm{TO}}^2$

| | | Phonon frequency($cm^{-1}$) | | | | | |
|---|---|---|---|---|---|---|---|
| Material | Mode | E=0 | E∥a-b | E∥c | $\sqrt{\Delta\omega^2}$ ($cm^{-1}$) | $\varepsilon_{\perp}^{lat}$ | $\varepsilon_{\parallel}^{lat}$ |
| *c*-BN | $T_2$ | 1027 | 1269 | 1269 | 745.4 | 2.38 | 2.38 |
| *w*-BN | $A_1$ | 1001 | 1001 | 1249 | 747 | 0 | 2.58 |
| | $E_1$ | 1041 | 1267 | 1041 | 722 | 2.16 | 0 |
| *h*-BN | $A_{2u}$ | 746 | 746 | 819 | 338 | 0 | 0.61 |
| | $E_{1u}$ | 1372 | 1610 | 1372 | 842 | 1.83 | 0 |

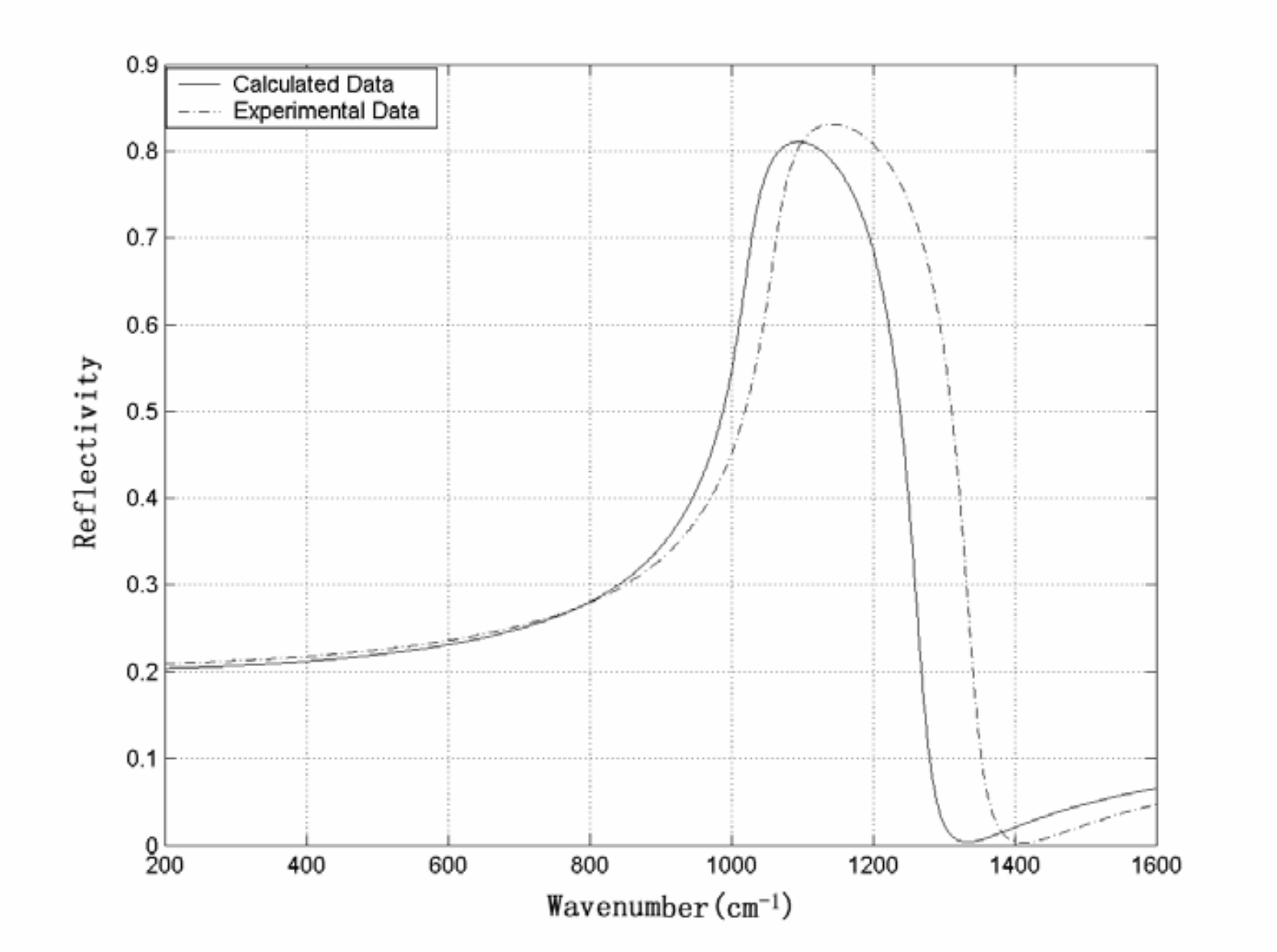

Fig.1. Frequency dependence of the infrared reflectance for *c*-BN. The dashdot line represents experimental spectrum. The calculated data are given by the solid line. The damping was chosen to be 3% of the frequency for each mode.

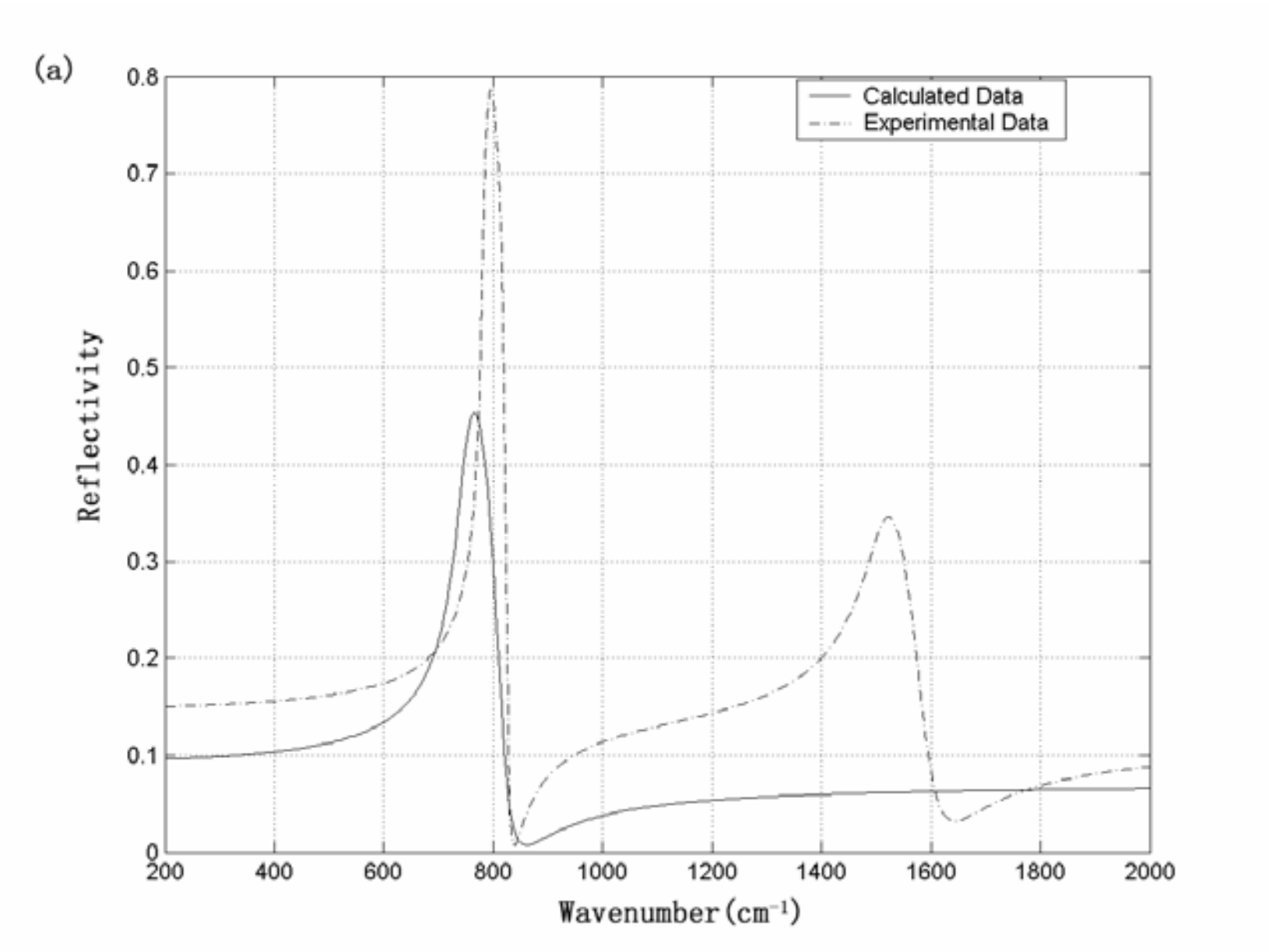

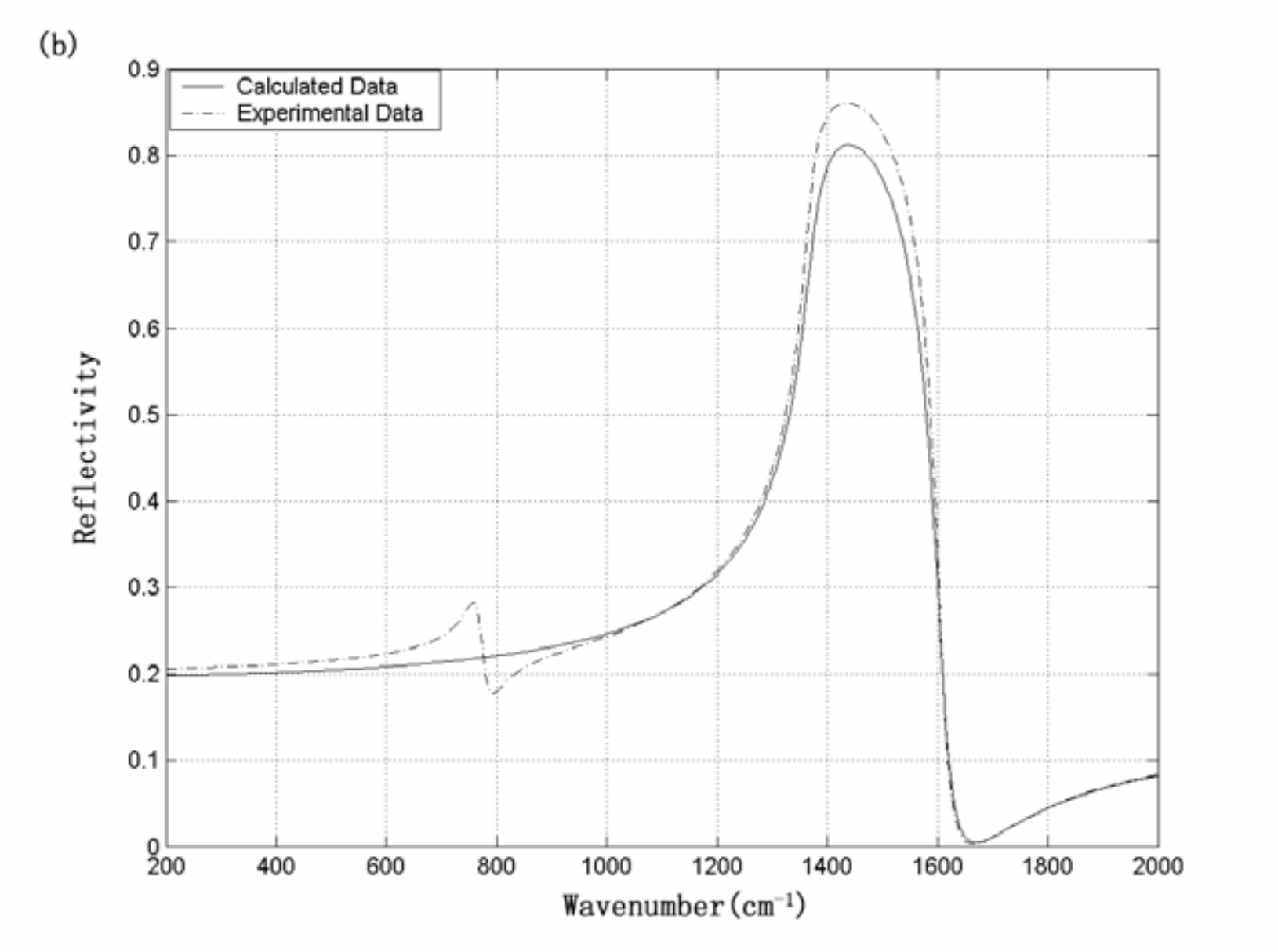

Fig. 2. Calculated (solid line) and experimental (dashdot line) reflectivity spectrum $R(\omega)$ for *h*-BN. The damping was chosen to be 3% of the frequency for each mode. (a) Polarization along the *c* axis. (b) Polarization in the *a*-*b* layer.

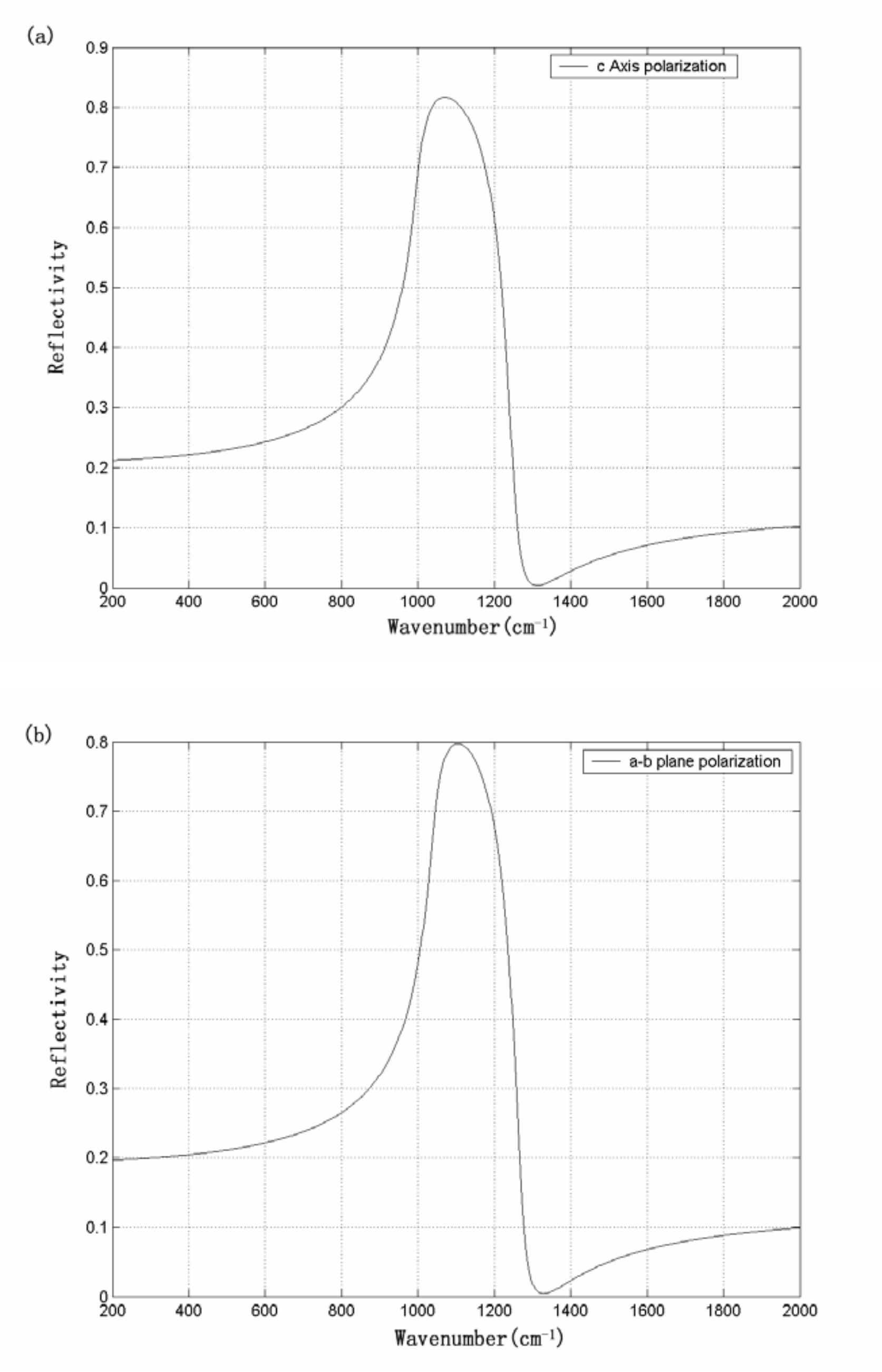

Fig. 3. Calculated reflectivity spectrum $R(\omega)$ for *w*-BN. The damping was chosen to be 3% of the frequency for each mode. (a) Polarization along the *c* axis. (b) Polarization in the *a*-*b* layer.